\documentclass[prl,aps,twocolumn,showpacs,floatfix,superscriptaddress,preprintnumbers]{revtex4-1}
\usepackage{epsfig}
\usepackage{psfrag}
\usepackage{bm}
\usepackage{tabularx}
\usepackage{amsmath,amssymb,bm,dcolumn}
\usepackage{ulem}
\usepackage{color}

\newcommand{\beq}{\begin{equation}}
\newcommand{\eeq}{\end{equation}}
\newcommand{\bea}{\begin{eqnarray}}
\newcommand{\eea}{\end{eqnarray}}

\begin{document}

\preprint{LA-UR-12-24539}

\title{Precise determination of the structure factor and contact in a unitary Fermi gas}

\author{Sascha Hoinka} 
\author{Marcus Lingham}
\author{Kristian Fenech}  
\author{Hui Hu}
\author{Chris J. Vale}
\affiliation{Centre for Atom Optics and Ultrafast Spectroscopy,
Swinburne University of Technology, Melbourne 3122, Australia}

\author{Joaqu\'{\i}n E. Drut}
\affiliation{Department of Physics and Astronomy, University of North Carolina, Chapel Hill, NC 27599-3255, USA}
\affiliation{Theoretical Division, Los Alamos National Laboratory, Los Alamos, New Mexico 87545, USA}

\author{Stefano Gandolfi}
\affiliation{Theoretical Division, Los Alamos National Laboratory, Los Alamos, New Mexico 87545, USA}

\date {\today}

\begin{abstract}
We present a high-precision determination of the universal contact parameter in a strongly interacting Fermi gas. In a trapped gas at unitarity we find the contact to be $3.06 \pm 0.08$ at a temperature of $0.08$ of the Fermi temperature in a harmonic trap.  The contact governs the high-momentum (short-range) properties of these systems and this low temperature measurement provides a new benchmark for the zero temperature homogeneous contact. The experimental measurement utilises Bragg spectroscopy to obtain the dynamic and static structure factors of ultracold Fermi gases at high momentum in the unitarity and molecular Bose-Einstein condensate (BEC) regimes. We have also performed quantum Monte Carlo calculations of the static properties, extending from the weakly coupled Bardeen-Cooper-Schrieffer (BCS) regime to the strongly coupled BEC case, which show agreement with experiment at the level of a few percent.
\end{abstract}


\maketitle


Ultracold atomic gases are unparalleled as a means to quantitatively probe strongly coupled 
systems which lie at the intersection of atomic~\cite{Giorgini:2008}, condensed-matter~\cite{AtomsReview1}, 
nuclear~\cite{Kaplan1998390,PhysRevC.77.032801} and
high-energy physics~\cite{Physics.2.88,Schafer:2009dj}.
In this context, two-component Fermi gases near a Feshbach resonance have particular 
significance as they are generally stable against inelastic decay. These {\it universal} 
quantum systems~\cite{PhysRevLett.92.090402,PhysRevA.63.043606,BraatenHammerReview} 
are characterized by strong coupling in the form of $s$-wave 
interactions of short range $r_0^{}$ and large scattering length $a$, such 
that the only scales left are thermodynamical: the density $n$ or chemical potential $\mu$, and temperature $T$, as for an ideal gas. This 
situation, in particular the unitarity limit $0 \leftarrow k^{}_F r^{}_0 \ll 1 \ll k^{}_F a \rightarrow \infty$, where $k^{}_F$ is the Fermi wavevector~\cite{ZwergerBook}, represents a major theoretical challenge as there are no small parameters. While no exact description exists, a variety of approximate techniques have been developed; however, these often give quite different predictions.

One of the key quantities characterizing these 
systems is the universal contact parameter $C$, introduced by 
Tan~\cite{Tan:2008a, Braaten:2008}. 
The contact derives from the short-range correlations in 
strongly interacting quantum gases and is the cornerstone of a 
number of exact relations describing properties such as the equation of state and dynamic response 
functions~\cite{ZhangLeggett,BraatenPlatter2,SonThompson,TaylorRanderia}.  
Evaluating these exact relations requires precise knowledge of $C$ itself, 
which is very challenging to compute with different calculations varying 
by as much as 10$\%$ \cite{Hu:2011,PhysRevA.85.053643}.

In this letter we provide a new experimental benchmark measurement, 
with error bars at the 3$\%$ level, for the contact at unitarity.  This is furnished by a precise determination of the dynamic and static structure factors 
using Bragg spectroscopy. In addition, we present new Quantum Monte Carlo 
(QMC) calculations accurate to the level of a few percent.  Our results indicate that theory and 
experiment are approaching a new level of convergence, showing that this difficult 
problem is becoming tractable.


{\it Experiments.}
The experiments presented here use a gas of $^6$Li atoms prepared in an
equal mixture of the $|F=1/2, m_F=\pm 1/2 \rangle$ spin states, evaporatively
cooled in a single-beam optical dipole trap.  Interactions are tuned
to the unitarity limit by setting the magnetic field to 833.0 G, near the pole of a broad Feshbach resonance.  Following evaporation, the
cloud of approximately $N/2 = (300 \pm 25)  \times 10^3$ atoms per spin state is loaded into a second optical dipole trap produced by a 10 W single frequency 1064 nm fibre laser, spatially filtered to produce a deep trap with large harmonic region.  We calibrate the atom number by imaging atom clouds with very high beam intensities~\cite{Reinaudi:2007,Esteve:2008} and verify this result with a precise measurement of the cloud size for a low temperature gas with weak attractive interactions where the modified Fermi radius can be calculated with high accuracy~\cite{Giorgini:2008}.  The temperature at unitarity in the final trap is $0.08 \pm 0.01 \, T_F^\text{HO}$ where $E_F^\text{HO} = k_B T_F^\text{HO} = (3N)^{1/3} \hbar \bar{\omega}$ is the Fermi energy in a harmonic trap, $\bar{\omega} = (\omega_x \omega_y \omega_z)^{1/3}$, $\omega^{}_x = \omega^{}_y = 2\pi \times 97$ Hz and $\omega^{}_z = 2\pi\times 24.5$ Hz.  We determine the temperature by fitting a Bold Diagramatic Monte Carlo prediction for the pressure equation of state~\cite{VanHoucke:2012} to the measured pressure obtained from one-dimensional (doubly integrated) density profiles~\cite{Ho:2010,Nascimbene:2010}.

Bragg scattering is performed as in previous 
work~\cite{Veeravalli:2008,Hoinka:2012}.  Two laser beams, detuned by
approximately 600 MHz from the nearest atomic transition, illuminate the
atom cloud, intersecting at an angle of $2 \theta = 84^\circ$.  This sets the probe wavevector $k = (4 \pi / \lambda) \sin{\theta}$ where $\lambda = 671$nm.  The beams are
derived from the same laser with two separate acousto-optic modulators,
driven by an amplified signal from a multichannel direct digital
synthesizer.  This allows us to precisely tune the relative frequency difference $\omega$ with an accuracy better than 1 Hz.

To obtain the dynamic response we measure the momentum imparted
to the cloud $\Delta P$ for a range of Bragg frequencies.  By imaging atoms in both
spin states with a short (850 $\mu$s) time delay between images we can
determine the centre of mass cloud displacement $\Delta X (\propto \Delta P)$~\cite{Veeravalli:2008} while remaining insensitive to
drifts in the initial cloud position.  We also carefully determine the
maximum Bragg laser intensities we can use at different Bragg frequencies and stay in the linear response
regime~\cite{Hoinka:2012}.  Scaling our data by the product of the
Bragg beam intensities allows us to combine data measured at different intensities into
a single spectrum and optimize the signal to noise.  This, along with averaging 10-15 points at each $\omega$, greatly
improves our measurement accuracy.
\begin{figure}
\begin{center}
\includegraphics[width=\columnwidth]{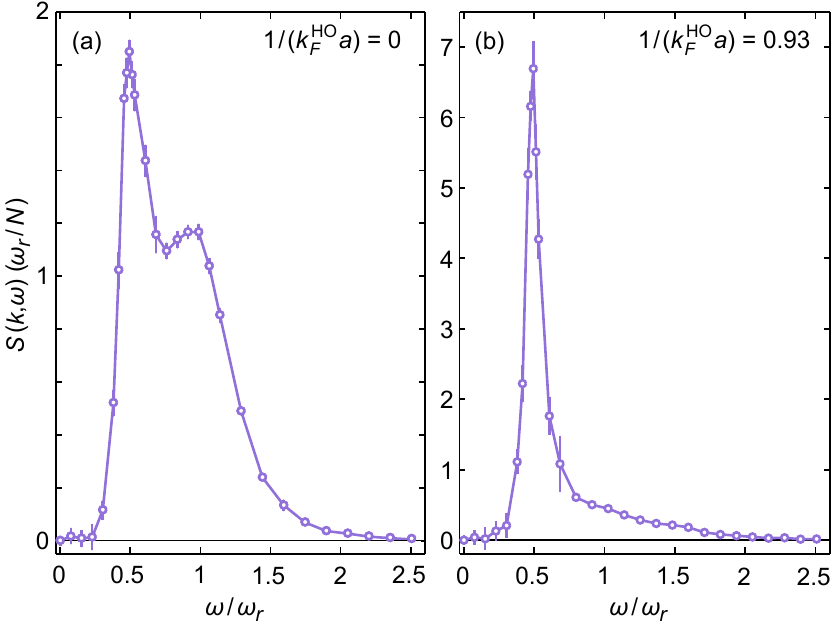}
\caption{(color online) Density-density response of a strongly interacting Fermi gas (a) at unitarity (833 G) and (b) at $1/(k^\text{HO}_F a) = 0.93$ (783 G), as a function of $\omega/\omega_r^{}$.  For the Bragg momentum $k=4.20k^\text{HO}_F$ these normalized spectra give the dynamic structure factor $S(k,\omega)$ in units of $\omega_r/N$. }
\label{fig:exp}
\end{center}
\end{figure}

Figure~\ref{fig:exp} shows Bragg spectra obtained (a) at unitarity 
and (b) at $1/(k^\text{HO}_F a)=0.93 $ where $\hbar k_F^\text{HO} \equiv (2 m E_F^\text{HO})^{1/2} = 2.97 \pm 0.05$ $\mu$m$^{-1}$ is the Fermi wavevector in the harmonic trap and $m$ is the $^6$Li atomic mass.  The uncertainty in $k_F^\text{HO}$ is dominated by the atom number uncertainty. These spectra contain a narrow peak at $\omega_r/2$ arising from the scattering of pairs and a broader feature centred around $\omega_r$ \cite{Hoinka:2012}.  Error bars are the statistical standard deviation of the data obtained at a particular frequency.  At the high momentum used here ($k = 4.20 k^\text{HO}_F$) the response is proportional to the dynamic structure factor $S(k,\omega)$~\cite{Combescot:2006,Veeravalli:2008}.  We scale each spectrum by its first energy-weighted moment making use of the $f$-sum rule for $S(k,\omega)$.  This means the quantity plotted
\begin{equation} 
\frac{\Delta X(\omega)}{\int
\omega \Delta X(\omega) d\omega} \equiv S(k,\omega) / \omega_r^{} 
\end{equation}
is the dynamic structure factor in units of $\omega_r^{} / N$ where $\omega_r = \hbar k^2 / (2m)$ is the recoil frequency.  The integral of these normalized spectra over $\omega$ gives the static structure factor~\cite{Kuhnle:2010} which we find to be 
$S(k=4.20k^\text{HO}_F) = 1.182 \pm 0.004$ at unitarity, and $1.50 \pm 0.02$ 
at $1/(k_F^\text{HO} a) = 0.93$.  The error bars account for the statistical uncertainties in the data. 

With this precise determination of $S(k)$ we can also obtain a new
measure of Tan's universal contact parameter $\mathcal I$ for a 
trapped gas at unitarity.  At this high momentum, the dimensionless contact can be found directly as~\cite{Combescot:2006,Hoinka:2012,KuhnleNJP:2011}
%
\begin{equation} 
\frac{\mathcal I}{Nk^\text{HO}_F} = \frac{4 k}{k^\text{HO}_F} \left( \frac{S(k) - 1}{1-4/(\pi k a)} \right)\,
\end{equation}

Using this expression we find ${\mathcal I}/({Nk^\text{HO}_F}) = 3.06 \pm 0.08$ at unitarity and $11.9 \pm 0.3$ at $1/(k_F^\text{HO}a)=0.93 \pm 0.02$.  The error bars include the uncertainty in $k_F$ as well as $S(k)$.  This is significantly more accurate than previously published data~\cite{Hu:2011,KuhnleNJP:2011} and sets a new benchmark for theoretical calculations. Our value at unitarity is some way below that obtained from measurements of the equation of state~\cite{Navon:2010} and from the frequency of collective oscillations~\cite{Yun:2011}; however, it is higher than found using radio-frequency spectroscopy~\cite{Stewart:2010,Sagi:2012} and photo-association data~\cite{Partridge:2005,Werner:2009}.  Comparing with different theoretical predictions our measurement is closest to, but slightly above, recent many-body $t$-matrix calculations \cite{Palestini:2010, Enss2011770}, yet lies below both a Nozi\`{e}res-Schmitt-Rink (NSR) calculation~\cite{Hu:2011} and zero temperature QMC results~\cite{Gandolfi:2011}.  At $1/(k_F^\text{HO}a)=0.93$ our measured contact $11.9 \pm 0.3$ is slightly below theory ($\sim 12.35$) possibly due to finite temperature.




{\it Quantum Monte Carlo calculations.}
We use QMC techniques on a system of 66 fermions to accurately compute the ground-state properties of strongly interacting fermions in the thermodynamic 
limit~\cite{Forbes:2011,Carlson:2011}.  We employ the same technique as in Ref.~\cite{Gandolfi:2011} to
calculate the dimensionless energy per particle $\xi = E/E_{FG}$ (the Bertsch parameter) 
as a function of $(k^{}_Fa)^{-1}$, where $E_{FG}$ is the corresponding energy of a 
non-interacting Fermi gas, $k^{}_F = (3\pi^2 n)^{1/3}$.  For each value of 
$(k^{}_Fa)^{-1}$ we perform a variational optimization of the many-body wave function, as described in Ref.~\cite{Carlson:2003,Gandolfi:2011}. The 
best variational ansatz is then used as a trial wave function for the projection in imaginary time.
The fixed-node approximation is used to control the sign problem, and
the accuracy of the energy and other properties depends on the quality
of the variational wave function. The fixed-node energy at unitarity
is within a few percent of the exact calculation of
Ref.~\cite{Carlson:2011}.

For each coupling strength we perform a different QMC calculation,
varying the effective range $r_e^{}$ of the two-body interaction to extrapolate to the $r_e\rightarrow0$ limit.
This is necessary because the energy per particle $\xi$ can be strongly dependent on $r_e^{}$,
for different $k^{}_F a$. For $(k^{}_F a)^{-1} > 0.2$, we find
that the slope of $E(r_e^{})$ is negative and drops quickly as $(k^{}_F a)^{-1}$ 
is increased \cite{Forbes:2012}. Extrapolating to the $r_e\rightarrow0$ limit is therefore crucial
in the BEC region, but much less important in the BCS regime.

\begin{figure}
\begin{center}
\includegraphics[width=\columnwidth]{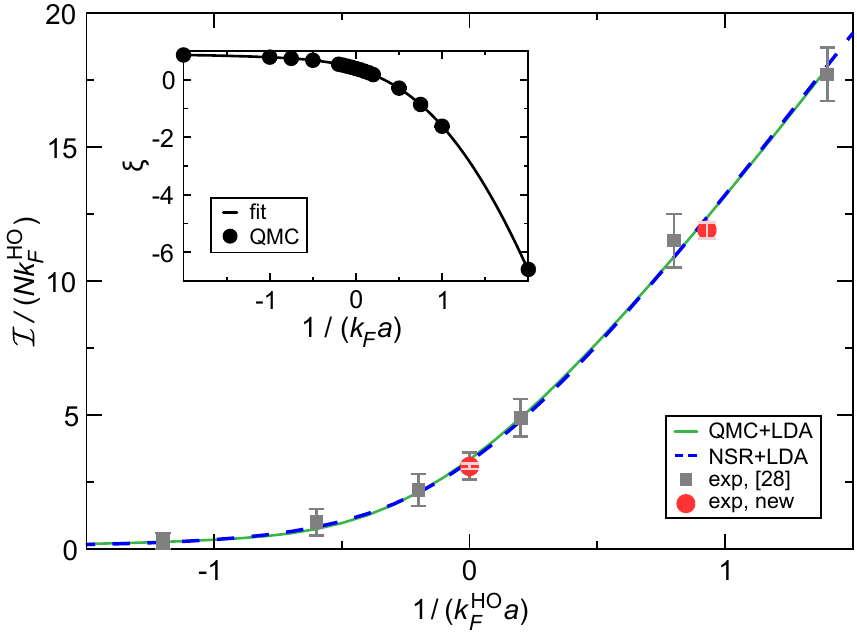}
\caption{(color online) Trap-averaged contact parameter as a function of $k^\text{HO}_F a$.  Green solid line is obtained from QMC data combined with the LDA and the black dotted line is a Nozi\`{e}res-Schmitt-Rink (NSR) calculation~\cite{Hu:2011}. Red circles are the new experimental points and the grey squares are experimental data from Ref.~\cite{KuhnleNJP:2011}. At unitarity, the experimental result is  ${\mathcal{I}}/{N k^\text{HO}_F} = 3.06 \pm 0.08$ compared with the QMC value: $3.336$ and NSR result: $3.26$~\cite{Hu:2011}. At $k^\text{HO}_F a=0.93$ the experiment yields $11.9 \pm 0.3$, while the QMC and NSR values are $\sim 12.35$~\cite{Hu:2011}.  Inset: Equation of state, points are QMC data and the solid line is a functional fit to these.  
}
\label{fig:contact_eos}
\end{center}
\end{figure}

The extrapolated value of $\xi$ as a function of $(k^{}_F a)^{-1}$ is shown in the inset of 
Fig.~\ref{fig:contact_eos}. We then use the adiabatic relation to calculate the contact for the
homogeneous system:
\begin{equation}
\label{eq:ContactFromEoS}
\frac{C}{Nk^{}_F}=-\frac{6\pi}{5}\frac{\partial \xi}{\partial (k^{}_F a)^{-1}} \,.
\end{equation}
Using Eq.~(3) we find the homogeneous contact to be 3.39 at unitarity and $\xi = 0.3899(4)$ which is approximately $4\%$ higher than a recent measurement~\cite{ZwierleinEtal}.

\begin{figure}
\begin{center}
\includegraphics[width=\columnwidth]{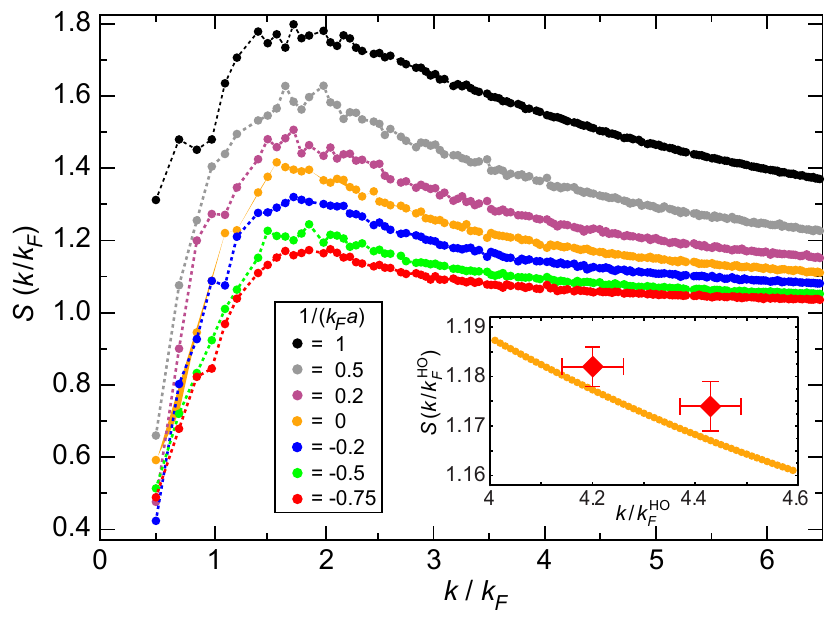}
\caption{(color online) Homogeneous static structure factor $S(k)$ as a function of $k/k_F^{}$, for various 
coupling strengths computed using QMC. In the inset we compare the trap-averaged calculation of $S(k)$ 
at unitarity with two new experimental results (red diamonds).  The measurements lie slightly above the QMC predictions.}
\label{fig:sofq}
\end{center}
\end{figure}

The static structure factor is computed via
\begin{equation}
S(k) = \langle\rho_k^\dagger  \rho_k^{}\rangle \,, 
\quad 
\rho_k = \sum_n \exp(i\boldsymbol k\cdot \boldsymbol r_n) \,.
\end{equation}
QMC results for the homogeneous $S(k)$ are shown in the main panel of Fig.~\ref{fig:sofq} for a wide range of coupling strengths
(calculated at $r_e\,k_F=0.056$). We have also calculated the spin-parallel component of $S(k)$, and
found $S_{\uparrow\uparrow}=S_{\downarrow\downarrow}=0.5$ within the error
bars for $k>4k_F$, in agreement with~\cite{Combescot:2006} and the experiments of~\cite{Hoinka:2012}.

{\it Local density approximation.}
Knowing the homogeneous contact and structure factor we can compute the trap-averaged values for these quantities using the local density approximation (LDA).  The overall chemical potential $\mu$ is related to the local chemical potential
$\mu(r)$ by $\mu = \mu(r) + V(r)$, where $V(r)$ is the trapping potential. At unitarity,
\beq
\mu(r) = \xi \varepsilon_F^{}(r) = \xi\frac{\hbar^2}{2m} \left( 3 \pi^2 n(r) \right)^{2/3}.
\eeq
Here $\varepsilon_F^{}(r)$ is the local Fermi energy, $n(r)$ is the density profile, and
we have assumed that $\xi$ does not depend on $r$ (valid in the unitary and free-gas limits);
more generally we have $\xi = \xi(k_F(r) a)$, where $k^{}_F(r) = (3 \pi^2 n(r))^{1/3}$ and $a$ is the scattering length.
For a harmonic trapping potential $V(r) = \frac{1}{2} m \omega^2 r^2$, 
the density profile is given by
\beq
n(r) = n(0)\left[ 1 - \frac{r^2}{R^2}\right]^{3/2}, \ \ \ \ \ \ R \equiv \left( \xi \varepsilon_F(0) \frac{2}{m \omega^2} \right)^{1/2},
\eeq
and $n(0)$ is determined by the total number of particles through the normalization
condition $\int d^3r \ n(r) = N$. Given $\xi$, $N$ and the frequency of the trap $\omega$, we completely 
determine the ground-state density profile. Within the LDA, the total contact in a trap is given by 
$\mathcal{I} = \int d^3 r \ \mathcal{I} (r)$, where
\beq
\frac{\mathcal{I} (r)}{n(r) k_F^{}(r)} = c_0 = \frac{C}{N k_F^{}},
\eeq
and $C$ is obtained from QMC data using Eq.~(\ref{eq:ContactFromEoS}). 
Using the density profile above, the trapped contact is
\beq
\frac{\mathcal{I}}{N k^\text{HO}_F} = \frac{256}{105 \pi}  c_0  \xi^{-1/4}.
\eeq
In Fig.~\ref{fig:contact_eos} we show the contact obtained from QMC 
simulation (solid green line), compared with the new experimental 
data presented in this letter (red circles) and the experimental 
data of Ref. \cite{KuhnleNJP:2011} (grey squares). 
The LDA for the static structure factor in a trap is
\beq
{\mathcal S} (k/k^\text{HO}_F) = \frac{1}{N}\int d^3 r\ n(r) S (k/k^{}_F(r)),
\eeq
where the dimensionless function $S(k/k^{}_F)$ was determined via QMC.  The inset of Fig.~\ref{fig:sofq} shows the trap-averaged QMC structure factor at unitarity and two new experimental points which lie just above the theory (both data points give the same value for the contact).


Away from unitarity $\xi = \xi(k^{}_F a)$, and
\beq
\frac{\mu}{\varepsilon_F^{}} = \xi(k^{}_F a) + \frac{1}{6 \pi k^{}_F a} \frac{C(k^{}_F a)}{N k_F^{}},
\eeq
which follows from $E = -PV + \mu N$ combined with $E = \frac{3}{2} PV - \frac{\hbar^2}{8m\pi a}C$.
As in the unitary case, the central density determines $\mu$, and the LDA equation $\mu = \mu(r) + V(r)$ determines the 
density profile, which is solved numerically as $k^{}_F a$ depends on $r$ through $k_F^{}(r)$. 

{\it Homogeneous zero-temperature contact.} 
While our measurements were performed on a trapped (inhomogeneous) cloud our result at unitarity also provides a constraint on the zero temperature homogeneous contact.  At finite temperatures the equation of state for the unitary Fermi gas is not known exactly, but, for $T \ll T_c$, where $T_c \, (\sim 0.2 \, T_F^{\text{HO}})$ is the superfluid transition temperature, trap averaged measurements are only weakly affected by the small population in the high temperature wings.  Several calculations of the temperature dependence of the trapped contact have recently been reported~\cite{Palestini:2010,Hu:2011, Enss2011770}, and, while these all vary significantly near $T_c$, at very low temperatures the predicted $T$-dependence is very similar.  Comparing the ratios ${\cal I} (T)/{\cal I} (0)$ one finds a weak $T$-dependence with a relative difference of order $1\%$ in the range $0 < T/T_F^{\text{HO}} < 0.08$ for the different models.  Thus, with knowledge of the trapped contact at a temperature $T \ll T_c$, we can extrapolate down to zero temperature and anticipate that systematics due to imprecise knowledge of the equation of state should be small.

Applying this extrapolation we obtain ${\mathcal I}_0/({Nk^\text{HO}_F}) = 3.15 \pm 0.09$ for the trapped contact as  $T \rightarrow 0$.  Using Eq.~(8) then gives $c_0 = 3.17 \pm 0.09$ for the zero temperature homogeneous contact density.  The increased error bar is due to the uncertainty in the extrapolation to $T = 0$.  We have used $\xi = 0.370 \pm 0.005$~\cite{ZwierleinEtal,Selim:2012} but note that the uncertainty in $\xi$ barely impacts the overall error as it appears in Eq.~(8) to the 1/4-th power.  An additional systematic arises from the fact that our measurement was performed at a magnetic field of 833.0 G which is not the exact field $B_0$ at which $|a| \rightarrow \infty$.  A recent determination found $B_0 = 832.18 \pm 0.08$ G~\cite{Selim:2012}, which, combined with the gradient of Fig.~2, would shift our result upwards by $\sim 2.5\%$.

In summary we have presented a high precision determination of the low temperature dynamic and static structure factors and contact of a strongly interacting Fermi gas.  These systems are an ideal testbed for validating different many-body calculations where exact predictions are not available.  Our measurements are now at a level that can discriminate between several of the established predictions and agreement with the latest QMC calculations is at the level of a few percent.  The measurement at unitarity also provides a new benchmark, with error bars at the $3\%$ level, for the $T \rightarrow 0$ limit of the homogeneous contact density which complements recent measurements at higher temperatures \cite{Sagi:2012}.  

{\it Acknowledgements.}
We would like to thank F. Werner for providing Bold Diagrammatic Monte Carlo data for the pressure equation of 
state and J. Carlson for useful discussions.  The work of J.E.D. and S.G. was supported by a grant from the Department
of Energy (DOE) under contracts DE-FC02-07ER41457 (UNEDF SciDAC),
and DE-AC52-06NA25396 (LANL). Computer time
was made available by Los Alamos Open Supercomputing, and by the National
Energy Research Scientific Computing Center (NERSC).



\begin{thebibliography}{99}

\bibitem{Giorgini:2008}
	S. Giorgini, L. P. Pitaevskii, and S. Stringari, 
	Rev.\ Mod.\ Phys. {\bf 80}, 1215 (2008);
	
\bibitem{AtomsReview1}
	I. Bloch, J. Dalibard, and W. Zwerger, 
	Rev.\ Mod.\ Phys. {\bf 80}, 885 (2008).

\bibitem{Kaplan1998390}
	D. B. Kaplan, M. J. Savage, and M. B. Wise, 
	Phys. Lett. B {\bf 424}, 390 (1998);
	Nucl. Phys. B {\bf 534}, 329 (1998).

\bibitem{PhysRevC.77.032801}
	A. Gezerlis and J. Carlson, 
  	Phys. Rev. C {\bf 77}, 032801 (2008).

\bibitem{Physics.2.88}
	T. Schaefer, 
	Physics {\bf 2}, 88 (2009).

\bibitem{Schafer:2009dj}
	T. Sch\"afer and D. Teaney, 
	Rept. Prog. Phys. {\bf 72}, 126001 (2009).

\bibitem{PhysRevLett.92.090402}
	T.-L. Ho, 
	Phys. Rev. Lett. {\bf 92}, 090402 (2004).

\bibitem{PhysRevA.63.043606}
	H. Heiselberg, 
    	Phys. Rev. A {\bf 63}, 043606 (2001).

\bibitem{BraatenHammerReview}
	E. Braaten and H.-W. Hammer, Phys. Rept. {\bf 428}, 259 (2006).

\bibitem{ZwergerBook}
   	W. Zwerger, {\it The BCS-BEC crossover and the Unitary Fermi Gas} (Springer, Berlin, 2011).

\bibitem{Tan:2008a}
   	S. Tan, 
   	Ann. Phys. {\bf 323}, 2952 (2008);
   	Ann. Phys. {\bf 323}, 2971 (2008);
   	Ann. Phys. {\bf 323}, 2987 (2008).

\bibitem{Braaten:2008}
    	E. Braaten and L. Platter, 
	Phys. Rev. Lett. {\bf 100}, 205301 (2008).

\bibitem{ZhangLeggett}
	 S.~Zhang and A.~J.~Leggett, 
	 Phys.\ Rev.\ A {\bf 77}, 033614 (2008).

\bibitem{BraatenPlatter2}
	 E.~Braaten, D.~Kang, and L.~Platter, 
	 Phys.\ Rev.\ Lett. {\bf 104}, 223004 (2010).
	 	
\bibitem{SonThompson}
	  D.~T.~Son and E.~G.~Thompson, 
	  Phys.\ Rev.\ A {\bf 81}, 063634 (2010).

\bibitem{TaylorRanderia}
	  E.~Taylor and M.~Randeria, 
	  Phys.\ Rev.\ A {\bf 81}, 053610 (2010).
	  
\bibitem{Hu:2011}
   	 H. Hu, X.-J. Liu, and P. D. Drummond,
	 New J. Phys. {\bf 13}, 035007 (2011).

\bibitem{PhysRevA.85.053643}
  	Y. Nishida, Phys. Rev. A {\bf 85}, 053643 (2012).
	
\bibitem{Reinaudi:2007} 
	G. Reinaudi, T. Lahaye, Z. Wang, and D. Gu\'{e}ry-Odelin, 
	Opt. Lett. {\bf 32}, 3143 (2007).

\bibitem{Esteve:2008}
	J. Est\`{e}ve, C. Gross, A. Weller, S. Giovanazzi, and M. K. Oberthaler,
	Nature {\bf 455}, 1216 (2008).

\bibitem{VanHoucke:2012}
	K. Van Houcke, F. Werner, E. Kozik, N. Prokof'ev, B. Svistunov, M. J. H. Ku, A. T. Sommer, L. W. Cheuk, A. Schirotzek, and M. W. Zwierlein,
	Nat. Phys. {\bf 8}, 366 (2012).

\bibitem{Ho:2010}
	T.-L. Ho and Q. Zhou,
	Nat. Phys. {\bf 6}, 131 (2010).
	
\bibitem{Nascimbene:2010}
	S. Nascimbene, N. Navon, K. J. Jiang, F. Chevy, and C. Salomon,
	Nature {\bf 463}, 1057 (2010).
	
\bibitem{Veeravalli:2008}
	 G. Veeravalli, E. Kuhnle, P. Dyke, and C. J. Vale,
	 Phys.\ Rev.\ Lett. {\bf 101}, 250403 (2008).

\bibitem{Hoinka:2012}
	S. Hoinka, M. Lingham, M. Delehaye, and C. J. Vale,
	Phys. Rev. Lett. {\bf 109}, 050403 (2012).

\bibitem{Combescot:2006}
    	R. Combescot, S. Giorgini, and S. Stringari,
    	Europhys. Lett. {\bf 75}, 695 (2006).

\bibitem{Kuhnle:2010}
  	E. D. Kuhnle, H. Hu, X.-J. Liu, P. Dyke, M. Mark, P. D. Drummond, P. Hannaford, and C. J. Vale,
  	Phys. Rev. Lett. {\bf 105}, 070402 (2010).
	 
\bibitem{KuhnleNJP:2011}
	E. D. Kuhnle, S. Hoinka, H. Hu, P. Dyke, P. Hannaford, and C. J. Vale,
	New J. Phys. {\bf 13}, 055010 (2011).

\bibitem{Navon:2010}
	N. Navon, S. Nascimb{\`e}ne, F. Chevy, and C. Salomon,
	Science {\bf 328}, 729 (2010).

\bibitem{Yun:2011}
   	Y. Li and S. Stringari,
	Phys. Rev. A {\bf 84}, 023628 (2011).
		 
\bibitem{Stewart:2010}
	J. T. Stewart, J. P. Gaebler, T. E. Drake, and D. S. Jin,
	Phys. Rev. Lett. {\bf 104}, 235301 (2010).
	
\bibitem{Sagi:2012}
	Y. Sagi, T. E. Drake, R. Paudel, and D. S. Jin,
	Phys. Rev. Lett. {\bf 109}, 220402 (2012).

\bibitem{Partridge:2005}
	G. B. Partridge, K. E. Strecker, R. I. Kamar, M. W. Jack, and R. G. Hulet,
	Phys.\ Rev.\ Lett. {\bf 95}, 020404 (2005).

\bibitem{Werner:2009}
   	F. Werner, L. Tarruell, and Y. Castin,
	Eur. Phys. J. B {\bf 68}, 401 (2009).

\bibitem{Palestini:2010}
	F. Palestini, A. Perali, P. Pieri, and G. C. Strinati,
	Phys. Rev. A {\bf 82}, 021605 (2010).

\bibitem{Enss2011770}
	T. Enss, R. Haussmann, and W. Zwerger,
	Ann. Phys. {\bf 326}, 770 (2011).

 \bibitem{Gandolfi:2011}
   	S. Gandolfi, K. E. Schmidt, and J. Carlson,
   	Phys. Rev. A {\bf 83}, 041601 (2011).
	
\bibitem{Carlson:2011}
	J. Carlson, S. Gandolfi, K. E. Schmidt, and S. Zhang,
	Phys. Rev. A {\bf 84}, 061602(R) (2011).

\bibitem{Forbes:2011}
 	M. M. Forbes, S. Gandolfi, and A. Gezerlis,
	Phys. Rev. Lett. {\bf 106}, 235303 (2011).

\bibitem{Carlson:2003}
    	J. Carlson,  S.-Y. Chang, V. R. Pandharipande, and K. E. Schmidt, 
   	 Phys. Rev. Lett. {\bf 91}, 050401 (2003).

\bibitem{Forbes:2012}
 	M. M. Forbes, S. Gandolfi, and A. Gezerlis,
	Phys. Rev. A {\bf 86}, 053603 (2012).
	
\bibitem{ZwierleinEtal}
	 M.~J.~H.~Ku, A.~T.~Sommer, L.~W.~Cheuk, and M.~W.~Zwierlein,
	 Science {\bf 335}, 563 (2012).

\bibitem{Selim:2012}
	G. Z\"{u}rn, T. Lompe, A. N. Wenz, S. Jochim, P. S. Julienne, and J. M. Hutson, arXiv:1211.1512 [cond-mat.quant-gas] (2012).
	


%
%
%
%
%
%
%


%
%
%
%
%
%
%


\end{thebibliography}
\end{document}